\documentclass[twocolumn,aps,prl,groupedaddress]{revtex4}
\usepackage{graphicx}
\usepackage{ulem}
\usepackage{color}
\usepackage{amsmath}
\usepackage{marvosym}
\newcommand{\be}{\begin{equation}}
\newcommand{\ee}{\end{equation}}
\newcommand{\bea}{\begin{eqnarray}}
\newcommand{\eea}{\end{eqnarray}}

\newcommand{\Mv}{{\bf M}}
\newcommand{\mv}{{\bf m}}
\newcommand{\Ev}{{\bf E}}
\newcommand{\jv}{{\bf j}}
\newcommand{\rv}{{\bf r}}
\newcommand{\qv}{{\bf q}}
\newcommand{\kv}{{\bf k}}
\newcommand{\zv}{{\bf z}}
\newcommand{\vv}{{\bf v}}

\newcommand{\Omegav}{{\bf \Omega}}
\newcommand{\la}{\langle}
\newcommand{\ra}{\rangle}

\renewcommand{\vec}[1]{{\bf #1}}
\def\nn{\nonumber\\}

\begin{document}

\title{Low-dissipation edge currents without edge states} 
\date{\today}
\author{Justin C. W. Song$^{1,2}$ and Giovanni Vignale$^{3,4}$}
\affiliation{$^1$ Division of Physics and Applied Physics, Nanyang Technological University, Singapore 637371}
\affiliation{$^2$ Institute of High Performance Computing, Agency for Science, Technology, \& Research, Singapore 138632}
\affiliation{$^3$ Department of Physics and Astronomy, University of Missouri, Columbia, Missouri 65211,~USA}
\affiliation{$^4$ Center for Advanced 2D materials, National University of Singapore, Singapore 117542}

\begin{abstract}
We show that bulk free carriers in topologically trivial multi-valley insulators with non-vanishing Berry curvature give rise to low-dissipation edge currents, which are squeezed within a distance of the order of the valley diffusion length from the edge.  This happens even in the absence of edge states [topological (gapless) or otherwise], and when the bulk equilibrium carrier concentration is thermally activated across the gap. 
Physically, the squeezed edge current arises from the spatially inhomogeneous orbital magnetization that develops from valley-density accumulation near the edge. While this current possesses neither topology nor symmetry protection and, as a result, is not immune to dissipation, in clean enough devices  it can mimic low-loss ballistic transport. 
\end{abstract} 
\maketitle

In bulk band insulators, carrier transport is exponentially activated, leading to a severely muted current response when an electric field is applied~\cite{ashcroftmermin}. However, this adage fails spectacularly in topological matter where gapped bulk bands, characterized by a non-trivial topology~\cite{tknn,wen}, 
support gapless edge states~\cite{halperin, buttiker, kanereview, wen}, which can carry dissipationlesss charge currents along the edges of the sample. As a result, such edge currents have become synonymous with topologically non-trivial bulk bands as expected from the principle of bulk-edge correspondence~\cite{hatsugai,wen,kanereview,streda2}. 

\begin{figure}[t!]
\includegraphics[width=\columnwidth]{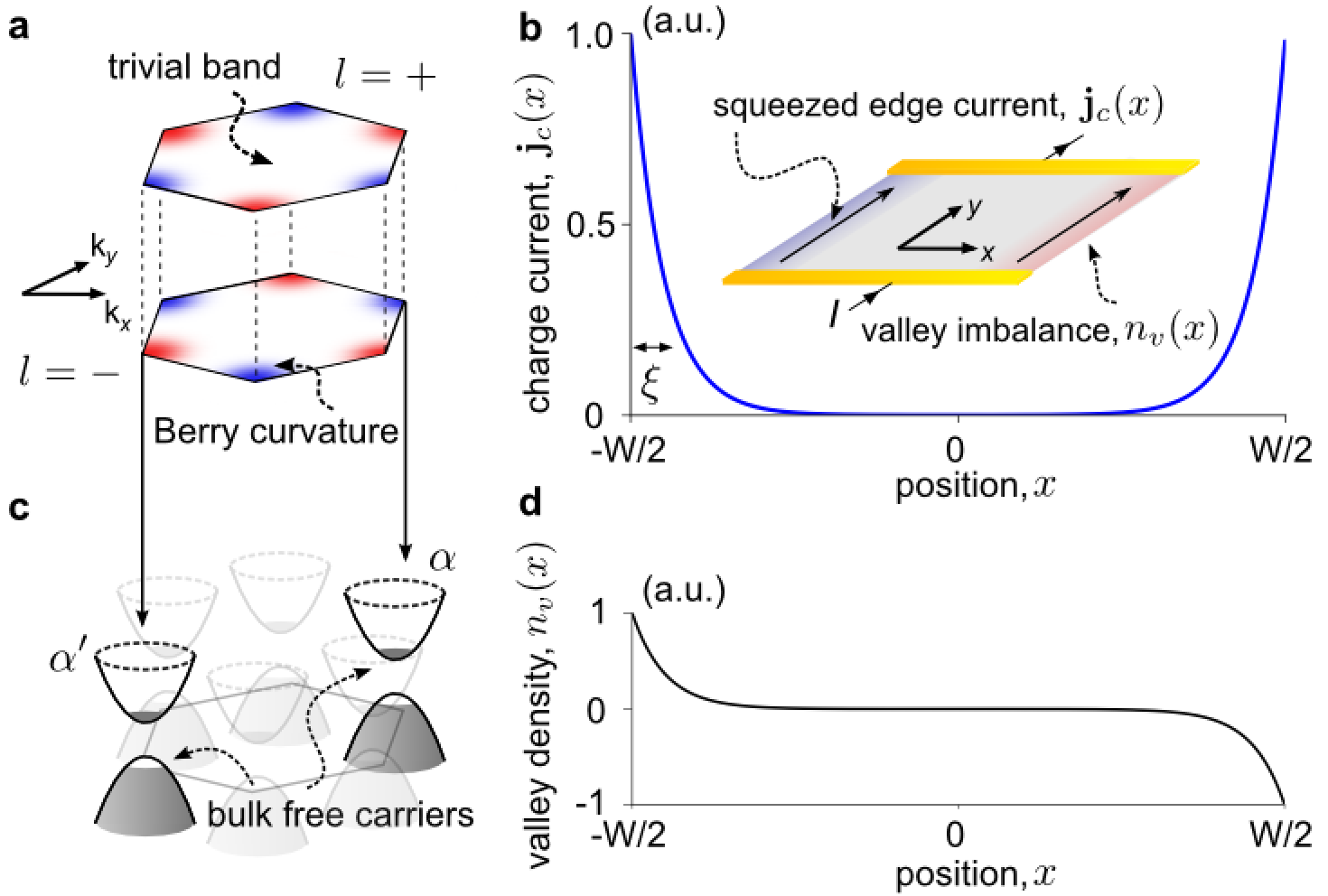}
\caption{{\bf Squeezed edge currents in a topologically trivial insulator.} {\bf a.} 
Berry curvature hot spots in topologically trivial insulator bands with zero net Berry flux over the entire Brillouin zone, e.g., (shown) Berry curvature, $\Omega_{l\alpha}$ hot spots for gapped graphene with broken inversion symmetry; $l=\pm$ are conduction and valence bands. {\bf b.} A {\it charged} squeezed edge current (SEC), $\vec j_c(\vec r)$, can flow along the sample edges [Eq.~(\ref{eq:squeezed})] even in a gapped finite sized device (inset) without 
edge states. 
{\bf c.} Carriers in highlighted bands at $\alpha, \alpha'$ experience opposite signs of Berry curvature and contrasting transport characteristics (see text). {\bf d.} Density imbalance between flavors/valleys can accumulate at sample edges over a width determined by the flavor/valley diffusion length, $\xi$, Eq.~(\ref{eq:nvprofile}). We have used $\Omega_{l\alpha}$ 
for a gapped Dirac material (see text) so that $\sigma_{H}^v >0$ and $\Theta_H<0$ [Eq.~(\ref{eq:jcjv}) and (\ref{eq:dxy})].} 
\label{fig1}
\end{figure}

Here we argue that in the presence of Bloch band Berry curvature, bulk free carriers in a multi-valley gapped insulator can conspire to produce a charge current  that is squeezed close to sample boundaries {\it in the absence of edge states} (Fig.~\ref{fig1}).  
The squeezed edge current (SEC) (Fig.~\ref{fig1}{\bf b}) has low (but finite) dissipation and occurs even when the equilibrium chemical potential is in the gap with a thermally activated bulk. 
As a result, SEC can act as a current conduit shunting the nominally insulating bulk
to produce unusual non-activated resistivity characteristics at low temperature. 

We expect SEC to naturally manifest in topologically trivial insulators possessing well-separated Bloch-band Berry curvature distributions~\cite{dixiao07} in the Brillouin zone (for e.g., in Fig.~\ref{fig1}), such that the total integrated curvature is zero. As such, these systems do {\it not} possess gapless topologically protected edge states.
Instead, the Berry curvature in each of the valleys enables valley Hall currents to be induced by an applied electric field and produce a valley density accumulation (of bulk carriers) near the edge of the sample, while the net charge density remains zero.  The valley density gradient perpendicular to the edge produces a charge current flowing along the edge. 
This induced charge current (transverse to the valley density gradient) can be viewed as an anomalous transverse diffusion of carriers, with off-diagonal diffusion constants of different signs in different valleys --- a characteristic of carriers possessing finite Berry curvature.
 
\begin{figure}[t!]
\includegraphics[width=\columnwidth]{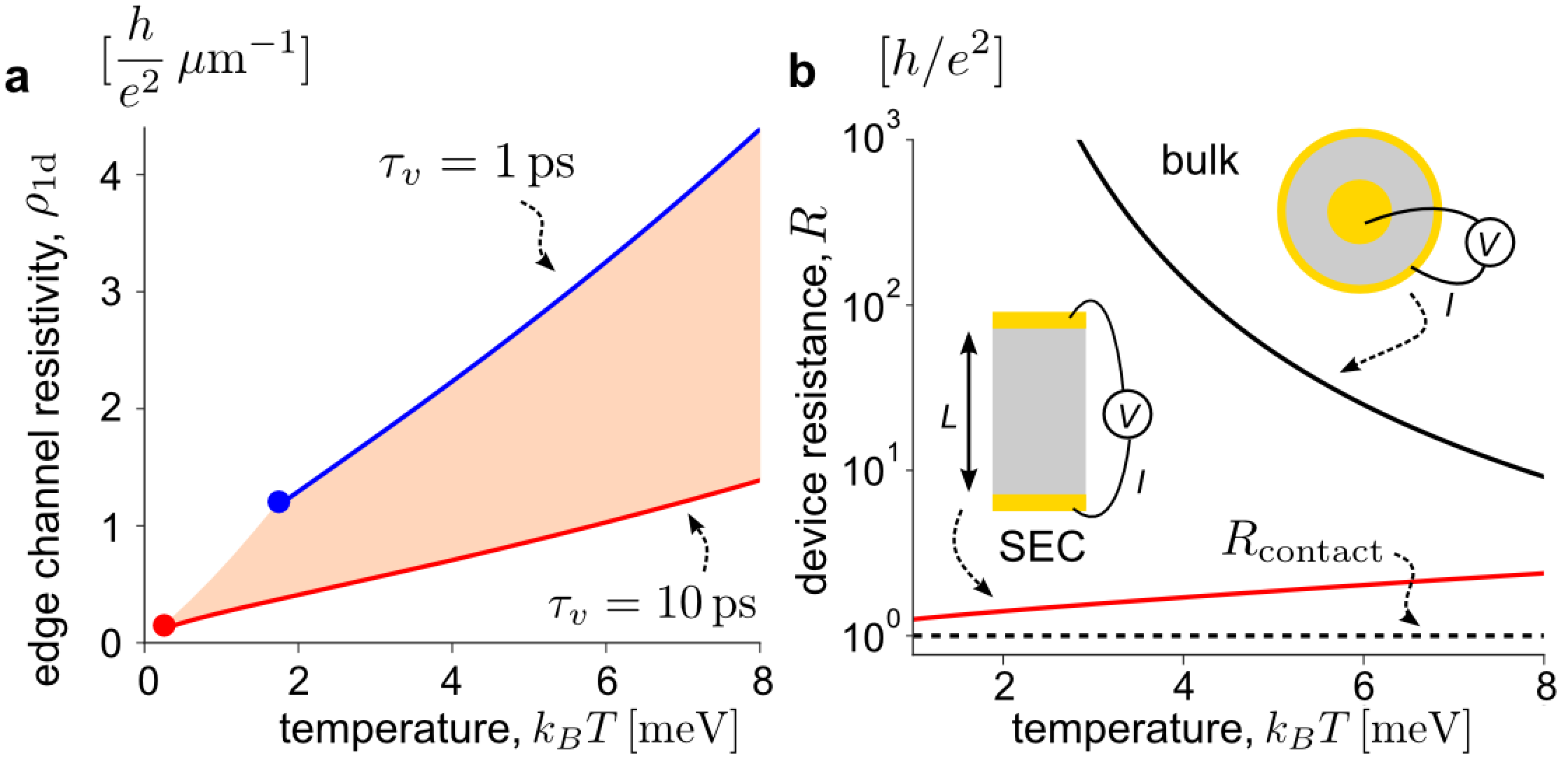}
\caption{{\bf Low dissipation squeezed edge channels.} {\bf a}. One-dimensional resistivity of a single squeezed edge current (SEC) channel along the edge of gapped graphene device [Eq.~(\ref{eq:diracsec})] shown for $\tau_v = 10\, {\rm ps}$ (red dashed) and $\tau_v = 1 \, {\rm ps}$ (blue dashed). $\tau_v$ in-between these two values occupy the shaded orange region. Red and blue dots indicate temperature $T_0$ above which the semi-classical treatment is valid for the respective $\tau_v$ [see Eq.~(\ref{eq:T0})]. {\bf b}. Device resistance for a Hall-bar device (red, $L=1\, \mu{\rm m}$ and $\tau_v = 10\, {\rm ps}$) and a Corbino device (black). For illustration we used parameters: $\Delta = 15\, {\rm meV}$, $\eta = 2 \, {\rm m}^2/{\rm Vs}$, and $\sigma_{H}^v = 2e^2/h$. Here we have taken a value of $R_{\rm contact} = h/e^2$. }
\label{fig2}
\end{figure}

SEC appears only in finite-sized sample (e.g., Hall-bar type geometries) and vanishes in the infinite bulk or when measurements exclude edge currents (e.g., Corbino geometries) (see Fig.~\ref{fig2}). While located close to sample boundaries, we emphasize that SEC arises from bulk carriers; it occurs 
in the absence of localized edge modes of either topological (gapless) edge state origin or from other sources (e.g., band bending~\cite{Fabrizio2016,Beukman2016}, gapped edge modes on rough boundaries~\cite{li11}). Instead, SEC is intimately tied to a current-induced bulk (out-of-plane) magnetization build-up at sample edges. 

While gapped graphene-type systems are not the only examples of this type of behavior, nevertheless, they present natural experimental targets due to their high quality, ease of manipulation, {\it lack} of topological gapless edge states, and clear observations of bulk valley Hall currents \cite{gorbachev14,tarucha15,zhang15}. Indeed, a recent experiment that infers edge-type currents in topologically trivial systems~\cite{geim} provide strong indications for SEC in gapped Dirac systems, see discussion below. 

{\it Inhomogeneous valley Hall currents -- } We begin by recalling that the position and velocity operators within a Bloch band ($l$) and valley ($\alpha$) are: 
\be\label{Position}
\hat \rv_{l\alpha} = i \frac{\partial}{\partial \kv}+\boldsymbol{\cal A}_{l\alpha}(\kv), \quad \hat \vv_{l\alpha} = \frac{1}{i\hbar}[\hat \rv_{l\alpha},\hat H],
\ee
where $\boldsymbol{\cal A}_{l\alpha}(\kv)=i\langle u_{l\alpha}(\kv)|\nabla_\kv u_{l\alpha}(\kv)\rangle$ is the Berry connection of the band and valley under consideration. We note that the band velocity reproduces the familiar $ \langle u_{l\alpha}(\kv)| \hat \vv_{l\alpha} |u_{l\alpha}(\kv)\rangle = \tfrac{d\epsilon_{l\alpha}(\kv)}{\hbar \partial\kv} - \hbar^{-1}e \Omegav_{l\alpha}(\kv) \times \Ev$, where  
$\Omegav_{l\alpha}(\kv)=\nabla_\kv \times\boldsymbol{\cal A}_{l\alpha}(\kv)$ is the Berry curvature, $\epsilon_{l\alpha}$ is the band energy, and $-e<0$ is the electron charge. For simplicity,  in what follows, we will consider only two Bloch bands separated by a gap: a valence band and a conduction band, with Berry curvature $\Omega_{l\alpha}\simeq \pm\lambda^2$ near the band extrema.  Here $\lambda$ plays the role of an effective ``Compton wavelength", inversely proportional to the gap at the band extrema. The above expressions are invariant under a gauge transformation that multiplies the Bloch wave function by a gauge-dependent phase.

We now construct the current density fluctuation operator at wavevector $\qv$ (for a single particle) as follows:
$\hat \jv_{l\alpha}(\qv )= -\frac{e}{2}\left(\hat \vv_{l\alpha} e^{-i\qv\cdot\hat\rv_{l\alpha}}+e^{-i\qv\cdot\hat\rv_{l\alpha}} \hat \vv_{l\alpha}\right)$. 
We will be interested in current distributions that are slowly varying on the scale of $\lambda$.  In this regime, we can expand $\hat \jv_{l\alpha}(\qv )$ 
to first order in $\qv$: 
\be\label{CurrentExpanded}
\hat \jv_{l\alpha}(\qv)= -e\hat{\vv}_{l\alpha}+\frac{i}{2}e\left[(\qv\cdot\hat\rv_{l\alpha})\hat\vv_{l\alpha}+\hat\vv_{l\alpha}(\qv\cdot\hat\rv_{l\alpha})\right].
\ee
While the first term in Eq.~(\ref{CurrentExpanded})
is the homogeneous current ($\qv=0$) see Eq.~(\ref{Position}), 
the second term only becomes relevant in an inhomogeneous system. Taking the latter's expectation value for state $| u_{l\alpha}(\kv)\ra$ yields a purely transverse current
$i \qv \times \mv_{l\alpha}(\kv)$, where $\mv_{l\alpha}(\kv)=-\frac{e}{4}\left(\hat \rv \times \hat\vv -\hat\vv \times \hat\rv\right)$ is the magnetic moment~\cite{Shi07}, see {\bf Supplementary Information (SI)}. 

The full physical current density 
in real space $\jv_{l\alpha}(\rv)$ proceeds directly from Eq.~(\ref{CurrentExpanded}). Performing an inverse Fourier transform, and averaging over a non-equilibrium state described by the inhomogeneous electron distribution function $f_{l\alpha}(\kv,\rv)$ yields
\bea\label{CurrentResult}
\jv_{l\alpha}(\rv) &=& \sum_{\kv} \left[-e\frac{\partial \epsilon_{l\alpha}(\vec k)}{\hbar \partial \vec k}  + \frac{e\Omegav_{l\alpha}(\kv)}{\hbar}\times e \Ev\right] f_{l\alpha}(\kv,\rv)\nonumber\\ &+& \sum_{\kv}  \frac{\partial  f_{l\alpha}(\kv,\rv)}{\partial \rv} \times \mv_{l\alpha}(\kv)\,. 
\eea
Here $f_{l\alpha}(\kv,\rv)$ is the distribution function.  
The first term of Eq.~(\ref{CurrentResult}) is the familiar homogeneous current (including a homogeneous Hall current driven by an electric field)~\cite{dixiao10}.  The second term is the current driven by
an electron density gradient, and exists even in the absence of direct mechanical forces (such as an applied electric field)~\cite{dixiao06,son12}. While {\it homogeneous} Hall currents driven by an electric field (second term in square brackets) can be sustained even in fully occupied bands at zero temperature~\cite{lensky15}, the {\it inhomogeneous} Hall current density [last term of Eq.~(\ref{CurrentResult})] is diffusive and requires a finite bulk band carrier density gradient. 

Na\"ively, one might expect that the transverse nature of the  
inhomogeneous Hall current does not contribute to charge transport since its divergence {\it deep in the bulk} vanishes~\cite{Cooper97}.  For example, smooth undulating variations in the bulk distribution function, $f_{l\alpha}(\kv,\rv)$, can cause circulating currents from the last term of Eq.~(\ref{CurrentResult}). While describing the local microscopic current, these circulate deep in the bulk and when summed across a large enough cross-sectional area, the net current vanishes. 

In contrast, the situation close to the edges of a sample is very different: the build up of density close to the edge does not result in circulating currents. Instead, as we will show, inhomogeneous Hall current freely flow as
charge currents. To ensure we capture the transport of charge we explicitly take a cross-section over the entire sample and integrate the net current flowing through it, see below. 

{\it Squeezed edge currents -- } In order to illustrate SEC, we will focus on a two-flavored Berry curvature hot spot system (indexed by $\alpha$), for example, that found in two-dimensional gapped graphene, where $\alpha = \{ K, K' \}$ for the two inequivalent valleys; $l=\{+,- \}$ for conduction and valence bands, see Fig.~\ref{fig1}. For brevity, in the following, we will drop the vector notation for Berry curvature since $\Omegav (\vec k) = \Omega (\vec k) \hat{{\vec z}}$ in two-dimensional systems.
Total {\it charge current} ($c$) is determined by $\jv_{c}\equiv \sum_{l,\alpha}  \jv_{l,\alpha}$ and the total {\it valley current} ($v$) is $\jv_{v}\equiv \sum_{l\alpha}  \alpha \jv_{l,\alpha}$ where $\alpha=1$ for $K$ and $\alpha=-1$ for $K'$. Similarly, we write charge and valley densities as 
$n_c \equiv \sum_{l,\alpha}  n_{l\alpha}$ 
and $n_v\equiv \sum_{l,\alpha}  \alpha n_{l\alpha}$; here $n_{l\alpha} (\vec r) = \sum_\vec k (-e) f_{l\alpha} (\vec k, \vec r)$ is the charge density in $l, \alpha$. 

Since $\Omega_{l\alpha}(\kv)$ 
changes sign in going from $\alpha =K$ to $\alpha = K'$, 
the flow of charge currents is particularly sensitive to the imbalance of distribution function between valleys. To see this, using Eq.~(\ref{CurrentResult}), we construct the total charge and valley currents in each band $l$ explicitly as
\bea
\jv_{c} && = -  \mathcal{D}_{xx}\boldsymbol{\nabla} n_c+ \sigma_{xx} \vec{E} -\Theta_H [(\boldsymbol{\nabla} n_v)\times \hat{\vec{z}}], \nonumber \\
\jv_{v} && =  -  \mathcal{D}_{xx}\boldsymbol{\nabla} n_v + [\sigma_{H}^v]  \hat{\vec{z}} \times \vec{E}
-\Theta_H [(\boldsymbol{\nabla} n_c )\times \hat{\vec{z}}],
\label{eq:jcjv}
\eea
where $\mathcal{D}_{xx}$ is the ordinary longitudinal diffusion constant of carriers within the bands, $\sigma_{xx}$ is the longitudinal conductivity, and $[\sigma_{H}^v] = (e^2/\hbar)\sum_{\vec k,l,\alpha} \Omega_{l\alpha} f_{l\alpha}^{(0)}(\vec k)$ is the valley Hall conductivity,  with $f_{l\alpha}^{(0)}(\vec k)$ is the Fermi-Dirac function $f_{l\alpha}^{(0)}(\vec k) = \{1+{\rm exp}\big[(\epsilon_{l}(\vec k) - \mu_{l\alpha})/(k_BT)\big] \}^{-1}$ with $\mu_{l \alpha}$ the (quasi-) chemical potential. Crucially, $\Theta_H$ captures current flow arising from an inhomogeneous distribution function in each of the valleys.   
This can be best seen by adopting a particle-hole symmetric two-band model such that  the energy of the conduction band $\epsilon_{+}>0$ is opposite to the energy of the valence band, $\epsilon_{-}=-\epsilon_+$, and both are independent of the valley index $\alpha$.  The magnetic moment 
is $\mv_{l\alpha}(\kv)=\frac{e}{\hbar} \epsilon_{l}(\kv) \Omega_{l\alpha}(\kv)\hat\zv$, and the inhomogeneous part of the current is therefore given, according to Eq.~(\ref{CurrentResult}), by $\jv_{l\alpha}=-\Theta_H^{l\alpha}\boldsymbol{\nabla} n_{l\alpha}\times\hat{\vec{z}}$, where 
\be
\Theta_H^{l\alpha}=\frac{\sum_{\vec k} \epsilon_{l}(\kv)\Omega_{l\alpha}(\kv)
\frac{\partial f_{l\alpha}^{(0)} (\kv)}{\partial \mu_{l\alpha}}} {\hbar \sum_{\vec k} \frac{\partial f_{l\alpha}^{(0)}(\kv)}{\partial \mu_{l\alpha}}}. 
\label{eq:dxy}
\ee
Now recall that the Berry curvature, 
$\Omega_{l\alpha}(\vec k)$, changes sign when either the band index or the valley index is switched: this implies that $\Theta_H^{+,\alpha} = \Theta_H^{-,\alpha} = \alpha \Theta_H$, where $\Theta_H\equiv \Theta_H^{l=+, \alpha=+1}$. Summing $\jv_{l\alpha}$ over $l$ and $\alpha$ gives the inhomogeneous charge current as written in Eq.~(\ref{eq:jcjv}).

When an electric field is applied along the sample, the bulk valley Hall effect produces a valley Hall current which must be cancelled by a valley density gradient perpendicular to the sample boundaries. This dramatically impacts {\it charge} transport characteristics. The profiles of density imbalance between valleys in each band 
$n_v(\vec r) $ obey the diffusion equation 
\be
\partial_t n_v(\vec r) - \mathcal{D}_{xx} \nabla^2 n_v(\vec r) + \frac{n_v(\vec r)}{\tau_v} = -\boldsymbol{\nabla} \cdot ([\sigma_{H}^v] \hat{\vec{z}} \times \vec{E}),
\label{eq:nvdiffusion}
\ee
where $\tau_v$  is the intervalley scattering time between valleys 
which captures the rate at which disparate parts (at $K$ and $K'$) of the Fermi surface equilibrate with each other. 
In the non-degenerate limit, the longitudinal diffusion can be estimated as $\mathcal{D}_{xx} = k_BT \eta /e$ where $\eta$ is the mobility; here we have used the same diffusion constant in both conduction and valence bands for simplicity. Different diffusion constants can be implemented with no qualitative change to the results below. 

Considering a long Hall bar, $L \gg W$, we treat $n_v(\vec r)$ and $\vec E(\vec r)$ as independent of $y$ along the bar; this reduces Eq.~(\ref{eq:nvdiffusion}), in the steady state, to a  one-dimensional differential equation, with the density jumping from a finite value to zero at $x=\pm W/2$. Further, by focusing on regions far away from contacts, we treat electric field as uniform. As a result, $n_v(\vec r)$ is driven only by delta-function sources at the boundaries $x = \pm W/2$: $ -\boldsymbol{\nabla} \cdot ([\sigma_{H}^v]  \hat{\vec{z}} \times \vec{E}) = - [\sigma_{H}^v]E \left[\delta (x- W/2)-\delta(x+W/2)\right]$, where $\vec E = E \hat{{\vec y}}$. We note that $\sigma_{H}^v$ is maximal when the chemical potential is in the gap~\cite{lensky15}.

The solution of the differential equation is found by elementary means to be
\be
n_v (x) =- \frac{[\sigma_{H}^v] E \tau_v}{\xi \cosh{(W/2\xi)}}\sinh{\left(\frac{x}{\xi}\right)},
\label{eq:nvprofile}
\ee
for $|x|\leq W/2$ and $0$ otherwise. Here $ \xi =\sqrt{\mathcal{D}_{xx}\tau_v}$ is the valley diffusion length. As shown in Fig.~\ref{fig1}{\bf d}, valley density accumulates at the edges. 

We emphasize that our diffusive treatment  
is valid only when the spatial profile of $n_v,\vec j_c$ is slowly varying on the scale of the Compton wavelength $\lambda = \hbar v/\epsilon $, the typical length scale of the wavepackets; for wavepackets close to the band edge in a gapped Dirac model, $\lambda_0 \approx \hbar v/\Delta$ (see Hamiltonian below), which is $\simeq 6\times 10^{-8}$ m for $v = 10^6 \,{\rm m}~{\rm s}^{-1}$ and half-gap size $\Delta = 10\, {\rm meV}$. 
The typical scale of $n_v(\vec r)$ variation is captured by the diffusion length $\xi$. As a result, we expect that our semi-classical diffusive picture holds as long as $\xi \gg \lambda$. Using the non-degenerate form of longitudinal diffusion constant $\mathcal{D}_{xx} = k_BT \eta/e$ we find this occurs for large enough temperatures
\be 
T \gg T_0, \quad k_BT_0 = \frac{e\lambda_0^2}{\eta\tau_v}. 
\label{eq:T0}
\ee  
Using a mobility $\eta = 1 \,{\rm m}^2 /({\rm Vs})$, $\tau_v = 10\, {\rm ps}$ we estimate $k_BT_0 \approx 0.4\, {\rm meV}$ ($T_0 \approx 5 \, {\rm K}$). Below this temperature scale (set by $T_0$), a fully quantum mechanical treatment is needed which is beyond the scope of the present work. In spite of this, the temperature regime $\Delta > T>T_0$ (in which our treatment is valid) defines a large and technologically important temperature regime.

Applying the inhomogeneous valley density profile in Eq.~(\ref{eq:nvprofile}) to Eq.~(\ref{eq:jcjv}) yields a charge current density flowing along the edge (see Fig.~\ref{fig1}) as
\be
\vec j_c^{\rm SEC}(\vec r) = j_c^{\rm SEC}(\vec r) \hat{\vec {y}}, \quad j_c^{\rm SEC} (\vec r) = 
\Theta_H 
 \partial_x n_v(\vec r).
\label{eq:squeezed}
\ee 

In the limit $\xi \ll W$, $\vec j_c^{\rm SEC} (\vec r)$ form squeezed quasi-one-dimensional channels flowing along the edges of the Hall bar. Crucially, Eq.~(\ref{eq:squeezed}) yields two squeezed current channels flowing in the {\it same} direction as shown in Fig.~\ref{fig1}; $\vec j_c^{\rm SEC} (\vec r)$ flows along $\vec E$. This demonstrates that the physical $\Theta_H$ current arising from the inhomogeneous electron distribution [see Eq.~(\ref{CurrentResult})] is {\it not} circulating, but contributes to total charge transport in the device.

Integrating the current density over one of these SEC channels and writing $E = V/L$ where $V$ is the voltage drop over length $L$ yields 
$ I_{\rm SEC}= \int_0^{W/2} j_c^{\rm SEC} (x) dx = -\Theta_H \sigma_{H}^v \tau_v V  /(\xi L) 
$. $I_{\rm SEC}$ constitutes a distinctly new parallel 
channel for current to flow in the Hall bar. We note that $-\Theta_H \sigma_{H}^v$ is positive, see Fig.~\ref{fig1}.
Adding the current flowing in the bulk, as well as accounting for contact resistance, we find the device resistance 
\be
\quad R^{-1} = R_{\rm bulk}^{-1} + R_{\rm SEC}^{-1}, \,\, R_{\rm SEC} = (\rho_{\rm 1d} L)+ R_{\rm contact}, 
\label{eq:fullresistance}
\ee
where $\rho_{\rm 1d} ={\xi}/{(|\Theta_H \sigma_{H}^v| \tau_v)}$, and $R_{\rm bulk}$ is the resistance of the bulk. 
Crucially, $\Theta_H, \sigma_{H}^v$ arise from the Berry curvature of the bands and exhibit a non-activated behavior in temperature, even when the chemical potential is in the gap. As we will see, this yields $\rho_{\rm 1d}$ that does not exponentially rise at low temperatures in stark contrast with $R_{\rm bulk}$ that exponentially rises at low temperatures. 

{\it SEC in gapped graphene -- }  We emphasize our above treatment is general and applicable to other multi-valley/flavor systems. For concreteness, however, and 
as an illustration of low-dissipation SEC in a non-topological insulator, we consider a gapped Dirac material, e.g. gapped graphene on hexagonal Boron Nitride, with Hamiltonian around each of the valleys as: $\mathcal{H}_{\alpha} = v\hbar(k_x \tau_x + \alpha k_y \tau_y) + \Delta \tau_z$ (where $\tau_{x,y,z}$ are Pauli matrices) and $\alpha = \pm 1$ for $K,K'$ valleys respectively.  The Berry curvature is $\Omega_{l\alpha} (\vec k)= -\frac{\alpha \lambda_0^2}{2} \frac{\Delta^3}{\epsilon_{l}^3(\vec k)}$.

In the non-degenerate limit $\mu_{l \alpha}, k_BT \ll \Delta$, we estimate $\sigma_{H}^v \approx 2 e^2/h$ for an almost fully filled band (accounting for spin degeneracy). Similarly, $\Theta_H$ can be estimated from Eq.~(\ref{eq:dxy}) in the same limit as
\be 
\Theta_H^{l\alpha} \approx \alpha \frac{\hbar v^2}{2\Delta}\mathcal{F}(\tilde\beta), \quad \mathcal{F}(\tilde\beta) = \Big[\frac{-\tilde\beta^2{\rm Ei}(-\tilde{\beta})}{(1+\tilde\beta){\rm exp}({-\tilde\beta})}\Big], 
\label{eq:dxydirac}
\ee
where $\tilde{\beta} = \Delta/k_BT$, ${\rm Ei} (x)= - \int_{-x}^\infty dt e^{-t}/t$ is the exponential integral, and we have approximated $(1+{\rm exp}[\tilde{\beta}])^{-1} \approx {\rm exp}[-\tilde{\beta}]$ for $\tilde{\beta} \gg 1$. Interestingly for small $T$, $\mathcal{F} \to 1$, reflecting the (band) geometrical origin of anomalous transverse diffusion. We note that $\sigma_{H}^v,\Theta_H$ do not vary significantly for $\vec E$-induced shifts in $\mu_{l \alpha} < \Delta$; sizeable valley imbalances along the edge can accumulate in the linear response regime.

Writing $\Theta_H  = \Theta_H^{l,(\alpha = +1)}$ [see Eq.~(\ref{eq:dxy})]
yields the resistivity of the quasi-1D channels along the sample edges 
\be
 \label{eq:diracsec}
\rho_{\rm 1d}^{(G)} (T) = \frac{\rho_0^{(G)}}{\tilde{\beta}^{1/2}\mathcal{F}(\tilde\beta)}, \quad \rho_0^{(G)} =  \frac{2\Delta^{3/2} (\eta/e)^{1/2}}{\hbar v^2 \tau_v^{1/2}|\sigma_{H}^v|}
\ee
where $ \rho_0^{(G)}$ is the characteristic 1D resistivity.

$\rho_0^{(G)}$ is non-universal and depends on the rate of relaxation of different parts of the Fermi surface at $K$ and $K'$ encoded in the intervalley scattering time $\tau_v$. While in a bulk homogeneous sample with few short-range impurities, intervalley scattering can be long (on the order of ten to several tens of picoseconds~\cite{gorbachev14,tarucha15,zhang15}), the value of $\tau_v$ can be reduced close to edges where surface roughness may enable enhanced intervalley scattering. Further, close to edges indirect scattering processes through flat/weakly-dispersive edge states can provide additional pathways for $K$ and $K'$ to relax~\cite{li10}. As an illustration, we use a range of $\tau_v \sim 1 - 10 \, {\rm ps}$ to estimate the value of $\rho_{\rm 1d}^{(G)} (T)$ yielding the shaded orange band of $\rho_{\rm 1d}^{(G)} (T)$ in Fig.~\ref{fig2}. 
 
Strikingly, 
even for relatively fast inter-valley scattering $\tau_v \sim 1 \, {\rm ps}$, 
$\rho_0^{(G)}$ can still take on small values $\rho_0^{(G)} \sim h/e^2  {\mu} {\rm m}^{-1}$, see Fig.~\ref{fig2}{\bf a} (red curve).
In contrast, the bulk resistance exponentially rises at low temperatures, $R_{\rm bulk} \propto {\rm exp} (\Delta/k_BT)$, where $\Delta$ is the half-gap size.  As a result, for small gap sizes of tens of meV, sufficiently short lengths, and low temperatures, SEC possess a very small resistivity [dominating $R^{-1}$ in Eq.~(\ref{eq:fullresistance})], and can mimic low-dissipation quasi-one-dimensional channel that shunts the bulk, see Fig.~\ref{fig2}{\bf b}. 

We note that in the low-temperature regime where $(\rho_{\rm 1d} L) \ll R_{\rm contact}$, Eq.~(\ref{eq:fullresistance}) is dominated by the contact resistance, see Fig.~\ref{fig2}b. Here we have chosen a simple $R_{\rm contact} = h/e^2$ as an illustration; other values of $R_{\rm contact}$ can be used with no qualitative changes to our conclusions. As a result of the low-dissipation in the SEC channel, current-voltage characteristics in a two-terminal geometry may display only very weakly 
$L$-dependent 
characteristics. Indeed, SEC may have been observed in recent transport experiments on gapped graphene-type structures where resistance saturated to a few resistance quanta at low temperatures in Hall-bar devices, and edge currents were identified using Josephson current spectroscopy~\cite{geim}. Further, recent Kerr-rotation microscopy in biased monolayer MoS$_2$ show magnetization accumulated along edges~\cite{shan16,shan17}, another signature of valley imbalance accumulation and SEC along the edge.

Bloch band quantum geometry can play a crucial role in charge transport of time-reversal invariant materials as epitomized by SEC that mimic ballistic edge channels without (spectral) edge states. 
SEC exhibits striking non-activated behavior even in nominally bulk insulating and topologically trivial devices. Additionally, SEC also mediates spin-free magneto-electric coupling, an unusual characteristic of these ``trivial'' insulators with non-vanishing Berry curvature; band geometry naturally interlaces charge and magnetization degrees of freedom even in a spin-orbit free system.

\vspace{2mm}
{\bf Acknowledgements} - We gratefully acknowledge useful conversations with Dima Pesin, Maksym Serbyn, and Valla Fatemi. This work was supported by the Singapore National Research Foundation (NRF) under NRF fellowship award NRF-NRFF2016-05 (J.C.W.S.). G.V. acknowledges generous support from the CA2DM, where this work was initiated, and from NSF Grant DMR-1406568.

\newpage
%

\setcounter{equation}{0}
\renewcommand{\theequation}{S-\arabic{equation}}
\makeatletter
\renewcommand\@biblabel[1]{S#1.}

\section{Supplementary Information for ``Low-dissipation edge currents without edge states''} 

\subsection{Covariant derivative and anomalous velocity}

As a warm-up, we briefly review the covariant derivative. Our starting point is the gauge invariant (physical) position operator in the Bloch representation
\be 
\hat \rv_{l\alpha} = i \frac{\partial}{\partial \kv}+\boldsymbol{\cal A}_{l\alpha}(\kv), 
\ee
where $\boldsymbol{\cal A}_{l\alpha}(\kv)=i\langle u_{l\alpha}(\kv)|\partial_\kv u_{l\alpha}(\kv)\rangle$ is the Berry connection of the band and valley under consideration. 
We note that $i\frac{\partial}{\partial \vec k} $ is the canonical (non-gauge-invariant) position operator in the momentum representation.

Crucially, different components of $\hat{\vec{r}}$ do not commute with each other. In particular, 
\be
 [\hat r_i,\hat r_j]  = i\Big(\partial_{k_i} [\mathcal{A}_{l\alpha}]_j-\partial_{k_j} [\mathcal{A}_{l\alpha}]_i\Big)\equiv i\varepsilon_{ijk}\Omega_k 
\ee
where the Berry curvature is 
\be
\Omega_i \equiv \varepsilon_{ijk}\partial_{k_j} [\mathcal{A}_{l\alpha}]_k\,. 
\ee

In the presence of an applied electric field, the Hamiltonian reads as $\hat H=\epsilon_n(\kv) - (-e) \Ev \cdot \hat \rv$. Here $-e < 0$ is the electron charge, and $\Ev$ is the electric field. Writing the velocity as $\hat \vv = \frac{1}{i\hbar}[\hat \rv,\hat H]$, we obtain 
\begin{align}
\la \hat v_i \ra & = \frac{1}{i\hbar} \la [r_i,\hat H] \ra  =  \frac{1}{\hbar}\frac{\partial \epsilon_n(\vec k )}{\partial_{k_i}}- \frac{i e}{\hbar} \la [\hat r_i,\hat r_j] \ra E_j \nn \\  
& = \frac{1}{\hbar}\frac{\partial \epsilon_n(\vec k )}{\partial_{k_i}}+ \frac{e}{\hbar}\varepsilon_{ijk} E_j \Omega_k,
\end{align} 
where the second term is the anomalous velocity. 

\subsection{Magnetic moment and inhomogeneous current density} 

In this section, we discuss the relationship between the magnetic moment and the inhomogeneous current density in Eq.~(\ref{CurrentExpanded}) of the main text. 

We begin by noting that the magnetic moment,
$
\hat \Mv =-\frac{e}{4}\left(\hat \rv \times \hat\vv -\hat\vv \times \hat\rv\right) 
$,  
can be re-expressed in component form as 
\be
\varepsilon_{ijk}M_k = \frac{i e}{2 \hbar}\left(\nabla_i \hat H \nabla_j-\hat H \nabla_i \nabla_j +\nabla_j  \nabla_i \hat H-\nabla_j \hat H \nabla_i\right),
\label{eq:Mexpanded}
\ee
where $\hat{H}$ is the Hamiltonian, and $\hat \rv = i\boldsymbol{\nabla}_\vec{k}$ with $\boldsymbol{\nabla_{\vec k}} = \partial_{\vec k} - i \boldsymbol{\mathcal{A}}_{l}(\vec k)$ the covariant derivative; $\nabla_j = [\boldsymbol{\nabla}_\vec{k}]_j$ is a shorthand. In obtaining the above expression, we have used $ \hat{v}_i = \hbar^{-1} [\nabla_i,\hat H]$ and the fact that $\nabla_j\nabla_k \hat H=\hat H \nabla_j\nabla_k$ when averaged in a Bloch state $|u_{l \alpha}\rangle$.
As a sanity check, we note that the expression for $\hat \Mv$ above here reduces to the well-known formula $ \la \hat \Mv \ra=
\mv_{l\alpha}(\kv) = \frac{i e}{2\hbar} \langle \nabla_\kv u_{l\alpha}(\kv)\vert(\epsilon_{l\alpha}(\kv)-\hat H)\times \vert\nabla_\kv u_{l\alpha}(\kv)\rangle$~\cite{dixiao10,Shi07}.

Similarly, we write the $q$-linear part of the current density operator in Eq.~(\ref{CurrentExpanded}) of the main text in component form as
\begin{align}
& i\frac{e}{2}q_j \left(\hat v_i \hat r_j+\hat r_j \hat v_i\right) \nonumber \\
& = -\frac{e}{2\hbar}q_j  \left(\nabla_i \hat H\nabla_j-\hat H \nabla_i\nabla_j+\nabla_j\nabla_i \hat H-\nabla_j \hat H\nabla_i\right)
\label{eq:Cexpanded}
\end{align}

Comparing Eq.~(\ref{eq:Cexpanded}) and Eq.~(\ref{eq:Mexpanded}) we obtain transverse current 
\be
i \frac{e}{2}q_j \la u_{la} |\hat v_i \hat r_j+\hat r_j \hat v_i | u_{la}\ra = i\la u_{la}| \varepsilon_{ijk}q_j \hat M_k |u_{la}\ra = i\vec q \times \vec m_{l \alpha} 
\ee
described in Eq.~(\ref{CurrentExpanded}) and Eq.~(\ref{CurrentResult}) of the main text.

\subsection{Alternative derivation of inhomogeneous current density: velocity matrix element}

In this section, we discuss an alternative algebraic derivation of the inhomogeneous current density by expanding the velocity matrix element 
at finite $\vec q$. For brevity, we will suppress the flavor index $\alpha$ leaving only the band index $l$ without loss of any generality. While less compact than the above discussion (using the magnetic moment operator), this alternative approach explicitly shows how the accumulation of geometric phases at finite $\vec q$ leads to the anomalous transverse diffusion. 

We proceed by considering the current dynamics in Bloch bands with a spatially varying out-of-equilibrium carrier density in the absence of an applied magnetic field.  The current density 
$ \mathbf{j}_l (\vec r) = e\sum_{\vec q} \vec{v}_\vec{q}^{(l)}  e^{i \vec q \cdot \vec r}$, can be expressed in terms of its Fourier harmonics as 
\be
 \vec{v}_\vec{q}^{(l)}= \sum_{\vec k} c^\dag_{\vec k_-, l} \la l, \vec k_- | \hat{\vec{v}} | l, \vec k_+ \ra c_{\vec k_+, l}, \quad \hbar \hat{\vec{v}} = \frac{\partial \hat{H}}{\partial \vec k},
 \label{eq:currentfull}
\ee
where $\hat{\vec{v}} $ is the velocity operator, $\hat{H}$ is the hamiltonian, $\vec k_\pm = \vec k + \vec q/2$, $c^\dag_{\vec k, l}$ is a creation operator for quasiparticles in band $l$ 
with corresponding (Bloch) wavefunction  $\la \vec r| c^\dag_{\vec k, l}| 0 \ra =\la \vec r | l, \vec k \ra e^{i\vec k \cdot \rv}$. The crystal wavefunctions $\la \vec r | n, \vec k \ra = u_{l,\vec k} (\vec r)$ are periodic over the unit cell.

As we now demonstrate, the phases accumulated by quasiparticles in the bands can play a crucial role in their transport, producing anomalous current flow when the carrier density is inhomogeneous. To illustrate this, we first note that the wavefunction $ \la \vec r | l, \vec k +\vec q \ra$ can be expanded, to leading order in $\vec q$, as
\be
\la  \vec r | l, \vec k \ra + \Big(\la \vec r | \frac{\partial u_{l,\vec k}}{\partial \vec k_{i}}\ra  - i\mathcal{A}_{l,i} (\vec k)\la \vec r | u_{l,\vec k}\ra   \Big)\vec q_i + \cdots,
\label{eq:wavefunctionexpand}
\ee
where we have expressed the expansion in component form, and $\mathcal{A}_{l,i} (\vec k) = \boldsymbol{\mathcal{A}}_{l} (\vec k) \cdot \hat{\vec x}_i$ is the $i^{\rm th}$ component of the Berry connection $\boldsymbol{\mathcal{A}}_l(\vec k)  = i\la u_{l,\vec k}| \partial_\vec{k} |u_{l,\vec k}  \ra$ (i.e. $\boldsymbol{\mathcal{A}}_{l}$ in the $\hat{\vec{x}}_i$ direction). Notice that the Taylor expansion in ${\vec k}$ is done using the covariant derivative,  $\boldsymbol{\nabla_{\vec k}} = \partial_{\vec k} - i \boldsymbol{\mathcal{A}}_{l}(\vec k)$: this is needed to ensure that
the calculated current is physical, i.e., invariant under a ``gauge transformation" of the crystal wave function, $u_{l,\vec k} (\vec r) \to e^{-i\chi({\vec k})}u_{l,\vec k} (\vec r)$.

Applying the expansion of the wavefunction at small $\vec q$ described in Eq.~(\ref{eq:wavefunctionexpand}) to the velocity matrix element, we obtain
\be
\la l, \vec k_- | \hat{\vec{v}}_i | l, \vec k_+ \ra = \la l, \vec k | \hat{\vec{v}}_i |l, \vec k \ra + [\boldsymbol{\mathcal{C}}^{(l)}_{ij} (\vec k) ](i\vec q_j) + \mathcal{O} (q^2), 
\label{eq:velocityexpand}
\ee
where $\hbar \la l, \vec k | \hat{\vec{v}}_i |l, \vec k \ra = \frac{\partial \epsilon_l(\vec k)}{\partial \vec k_i} $
is the group velocity, and 
\begin{widetext}
\begin{align}
[\boldsymbol{\mathcal{C}}^{(l)}_{ij} (\vec k) ](i\vec q_j) & = \Bigg[\big\la \frac{\partial u_{l,\vec k}}{\partial \vec k_j}\big| \hat{\vec{v}}_i \big| u_{l,\vec k} \big\ra -  \big\la u_{l,\vec k} \big| \hat{\vec{v}}_i \big| \frac{\partial u_{l,\vec k}}{\partial \vec k_j}\big\ra\Bigg]  \frac{\vec q_j}{2} -  2i \la u_{l,\vec k}| \hat{\vec{v}}_i |u_{l,\vec k} \ra \mathcal{A}_j \frac{\vec q_j}{2} \nonumber \\
& = \sum_m\Bigg[\big\la \frac{\partial u_{l,\vec k}}{\partial \vec k_j}\big|u_{m,\vec k} \ra\la u_{m,\vec k} | \hat{\vec{v}} _i\big| u_{l,\vec k} \big\ra -  \big\la u_{l,\vec k} \big| \hat{\vec{v}}_i \big| u_{m,\vec k} \ra\la u_{m,\vec k} | \frac{\partial u_{l,\vec k}}{\partial \vec k_j}\big\ra\Bigg]  \frac{\vec q_j}{2} - 2i \la u_{l,\vec k}| \hat{\vec{v}}_i |u_{l,\vec k} \ra \mathcal{A}_j \frac{\vec q_j}{2}. 
\label{eq:v1}
\end{align}
\end{widetext}
In the last line we have inserted the resolution of the identity $\sum_m |u_{m,\vec k} \ra \la u_{m,\vec k} |= 1$ into the terms of the square parentheses.

In order to proceed, we note that when $m=l$, the square parentheses cancel with the last term since $\la u_{l,\vec k} | \frac{\partial u_{l,\vec k}}{\partial \vec k_j} \ra = -i \mathcal{A}_j$. As a result, only terms with $m\neq l$ remain in Eq.~(\ref{eq:v1}). Using the identity for the interband matrix element
\be
\hbar \la u_{l,\vec k} |  \hat{\vec{v}}_i | u_{m,\vec k} \ra = \la u_{l,\vec k}|  \frac{\partial u_{m,\vec k}}{\partial k_i}\ra [\epsilon_l(\kv) - \epsilon_{m}(\kv)], \quad l \neq m, 
\ee
where  $\epsilon_l(\vec k)$ is the quasiparticle energy in band $l$, yields 
\begin{equation}
\boldsymbol{\mathcal{C}}^{(l)}_{ij}  =\frac{i}{2\hbar}\sum_{m \neq l} \la \frac{\partial u_{l,\vec k}}{\partial k_i}| u_{m,\vec k}\ra [\epsilon_l(\kv) - \epsilon_{m}(\kv)]\la  u_{m,\vec k}| \frac{\partial u_{l,\vec k}}{\partial k_j}\ra - c.c.
\label{eq:c}
\end{equation}

Comparing this with the well known expression for the magnetic moment~\cite{dixiao10,Shi07}
\be
\la \hat \Mv \ra=\frac{i e}{2\hbar} \langle \nabla_\kv u_{l\alpha}(\kv)\left\vert(\epsilon_{l\alpha}(\kv)-\hat H)\times \right\vert\nabla_\kv u_{l\alpha}(\kv)\rangle
\ee
 yields Eq.~(\ref{CurrentResult}) of the main text. 

\subsection{Estimate of characteristic SEC resistivity}

In the following we give a simple estimate of the characteristic SEC resistivity. Recalling Eq.~(\ref{eq:diracsec}) of the main text, we have the resistivity of the SEC channel 
\be
\rho_{\rm 1d}^{(G)} (T) = \frac{\rho_0^{(G)}}{\tilde{\beta}^{1/2}\mathcal{F}(\tilde\beta)}, \quad \rho_0^{(G)} =  \frac{2\Delta^{3/2} (\eta/e)^{1/2}}{\hbar v^2 \tau_v^{1/2}|\sigma_{H}^v|}, 
\ee
where $ \rho_0^{(G)}$ is the characteristic 1D resistivity and can be estimated as
\be
\rho_0^{(G)} = 0.48 \frac{(\Delta [{\rm meV}]/10)^{3/2} (\eta [{\rm m}^2/{\rm Vs}])^{1/2}}{(\tau_v [{\rm ps}]/10)^{1/2}} \left[\frac{h}{e^2} {\mu} {\rm m}^{-1}\right],
\ee
where we have used $v= 10^6 \, {\rm m}/{\rm s}$, and taken $|\sigma_{H}^v| = 2 e^2/h$.


\begin{thebibliography}{99}

\bibitem{ashcroftmermin} N. W. Ashcroft, and N.D. Mermin, {\it Solid State Physics}, Brooks/Cole Cengage Learning, California, (1976).


\bibitem{tknn} D. J. Thouless, M. Kohmoto, M. P. Nightingale, and M. den Nijs, ``Quantized Hall conductance in a two-dimensional periodic potential '' Physical Review Letters {\bf 49} 405 (1982)
\bibitem{wen} X. G. Wen, {\it Quantum field theory of many-body systems: from the origin of sound to an origin of light and electrons}. Oxford University Press (2004).

\bibitem{halperin} B. I. Halperin, ``Quantized Hall conductance, current-carrying edge states, and the existence of extended states in a two-dimensional disordered potential'' Physical Review B {\bf 25} 2185 (1982). 

\bibitem{buttiker} M. B\"uttiker, ``Absence of backscattering in the quantum Hall effect in multiprobe conductors'' Physical Review B {\bf 38} 9357 (1988)

\bibitem{kanereview} M. Z. Hasan, and C. L. Kane. ``Colloquium: topological insulators '' Reviews of Modern Physics 82.4 (2010): 3045.

\bibitem{hatsugai} Y. Hatsugai, ``Chern number and edge states in the integer quantum Hall effect '' Physical Review Letters {\bf 71} 3697 (1993). 

\bibitem{streda2}  A. H. Macdonald, P Streda, ``Quantized Hall effect and edge currents'', Physical Review B {\bf 29}, 1616 (1984). 

\bibitem{dixiao07} D. Xiao, W. Yao, and Q. Niu, ``Valley-contrasting physics in graphene: magnetic moment and topological transport." Physical Review Letters {\bf 99} 236809 (2007)

\bibitem{Fabrizio2016} N. Fabrizio, et al., ``Edge transport in the trivial phase of InAs/GaSb'' New Journal of Physics {\bf 18} 083005 (2016).

\bibitem{Beukman2016} A. J. A. Beukman, ``Topology in two-dimensional systems'' section 7.3, PhD Thesis TU Delft (2016). 


\bibitem{li11} Li, Jian, et al. "Topological origin of subgap conductance in insulating bilayer graphene." Nature Physics 7.1 (2011): 38.


\bibitem{gorbachev14} R. V. Gorbachev, et al., ``Detecting topological currents in graphene superlattices'' Science {\bf 346} 448-451 (2014)

\bibitem{tarucha15} Y. Shimazaki, et al., ``Generation and detection of pure valley current by electrically induced Berry curvature in bilayer graphene'' Nature Physics {\bf 11} 1032-1036 (2015)

\bibitem{zhang15} M. Sui, et al., ``Gate-tunable topological valley transport in bilayer graphene'', Nature Physics {\bf 11} 1027-1031 (2015).

\bibitem{geim} M. J. Zhu, et al., ``Edge currents shunt the insulating bulk in gapped graphene'',  Nature Communications {\bf 8} 14552 (2017).

\bibitem{Shi07}J. Shi, G. Vignale, Di Xiao, and Q. Niu, ``Quantum theory of orbital magnetization and its generalization to interacting systems'', Physical Review Letters {\bf 99}, 197202 (2007).

\bibitem{dixiao10} Xiao, Di, Ming-Che Chang, and Qian Niu, ``Berry phase effects on electronic properties'',  Reviews of modern physics {\bf 82} 1959 (2010)

\bibitem{dixiao06} D. Xiao., Y. Yao, Z. Fang, and Q. Niu ,``Berry-phase effect in anomalous thermoelectric transport'',  Physical Review Letters, {\bf 97} 026603 (2006).

\bibitem{son12} D. T. Son, and N. Yamamoto, ``Berry curvature, triangle anomalies, and the chiral magnetic effect in Fermi liquids'' Physical Review Letters {\bf 109} 181602 (2012). 

\bibitem{lensky15} Y. D. Lensky, J. C. W. Song, P. Samutpraphoot, L. S. Levitov, ``Topological valley currents in gapped Dirac materials'' Physical Review Letters {\bf 114} 25 (2015). 

\bibitem{Cooper97} N. R. Cooper, B. I Halperin, and I. M. Ruzin, ``Thermoelectric response of an interacting two-dimensional electron gas in a quantizing magnetic field'', Physical Review B {\bf 55}, 2344 (1997).

\bibitem{li10} L. Jian, A. F. Morpurgo, M. B\"uttiker, and I. Martin, ``Marginality of bulk-edge correspondence for single-valley Hamiltonians." Physical Review B {\bf 82} 245404 (2010).

\bibitem{shan16} J. Lee, K.F.  Mak, and J. Shan. ``Electrical control of the valley Hall effect in bilayer MoS 2 transistors.'' Nature nanotechnology {\bf 11} 421 (2016).
\bibitem{shan17} J. Lee, Z. Wang, H. Xie, K. F.  Mak, and J. Shan, ``Valley magnetoelectricity in single-layer MoS$_2$''' Nature materials, {\bf 16} 887 (2017).



\bibitem{supplement} See {\bf Supplementary Information} for a discussion of anomalous velocity and Berry connection, a detailed derivation of anomalous transverse diffusion via the orbital magnetic moment operator, an alternative analytic derivation via finite-$\vec q$ expansion of the velocity matrix element, and a discussion of the values of $\rho^{(0)}_G$. 

\end{thebibliography}
\end{document}